\documentclass{article}

\usepackage{natbib} % has a nice set of citation styles and commands
\usepackage{mathtools} % amsmath with fixes and additions
\usepackage{booktabs} % commands to create good-looking tables
\usepackage{tikz} % nice language for creating drawings and diagrams
\usepackage{graphicx}
\usepackage{algorithm2e}
  \SetKwInOut{Input}{Input}
  \SetKwInOut{Output}{Output}
  \usepackage{adjustbox}
\newtheorem{definition}{Definition}[section]

\usepackage{amsmath}
\usepackage{hyperref}
\hypersetup{colorlinks,allcolors=black}

\newcommand{\footremember}[2]{%
  \footnote{#2}
    \newcounter{#1}
    \setcounter{#1}{\value{footnote}}%
}
\newcommand{\footrecall}[1]{%
    \footnotemark[\value{#1}]%
} 
\usepackage{url}

\usepackage{graphicx}
\usepackage{float}
\usepackage{caption}
\usepackage{subcaption}
\usepackage{algorithm2e}
\usepackage{xcolor}
\usepackage{amsmath}
\usepackage{amssymb}

\title{Scalable Model Selection for Staged Trees: Mean-posterior Clustering and Binary Trees}

\author{Peter Strong \footremember{complexity}{Centre for Complexity Science, University of Warwick, Coventry CV4 7AL, UK}\footremember{turing}{The Alan Turing Institute, British Library, 96 Euston Road, London NW1 2DB, UK} \\ \href{mailto:P.R.Strong@warwick.ac.uk}{P.R.Strong@warwick.ac.uk} 
\and Jim Q Smith \footremember{stats}{Department of Statistics, University of Warwick, Coventry CV4 7AL, UK}\footrecall{turing} \\ \href{mailto:J.Q.Smith@warwick.ac.uk}{J.Q.Smith@warwick.ac.uk}} 
\begin{document}
\maketitle

\begin{abstract}
Several structure-learning algorithms for staged trees, asymmetric extensions of Bayesian networks, have been proposed. However, these either do not scale efficiently as the number of variables considered increases, \textit{a priori} restrict the set of models, or they do not find comparable models to existing methods. Here, we define an alternative algorithm based on a totally ordered hyperstage. We demonstrate how it can be used to obtain a quadratically-scaling structural learning algorithm for staged trees that restricts the model space \textit{a-posteriori}. Through comparative analysis, we show that through the ordering provided by the mean posterior distributions, we can outperform existing methods in both computational time and model score. This method also enables us to learn more complex relationships than existing model selection techniques by expanding the model space and illustrates how this can embellish inferences in a real study. 

\end{abstract}

\section{Introduction}
Probabilistic graphical models represent various factorisations of a probability distribution and, hence, conditional independencies. The graphical structure representing the factorisations can be elicited from expert judgement, but it is commonly learned using structure-learning algorithms. One such well-developed method is the Bayesian Network (BN). BNs represent symmetrical conditional independence statements between variables in a directed acyclic graph. 

Staged trees generalise finite discrete BNs in two ways. Firstly, they can represent more complex independence statements such as context-specific independence statements. Secondly, they can represent events that unfold in an asymmetric way. These extra properties make the class of staged trees widely applicable and they have been used in a number of domains including policing \cite{policing}, migration \cite{strong2021bayesian} and systems reliability \cite{systemsrel}.

The conditional independencies in a staged tree are represented through the use of colour on the non-leaf nodes -- called situations -- of the event tree. Model selection for staged trees is the task of exploring different partitions of the set of situations to find the \textit{maximum a posteriori} (MAP). However, the number of partitions of a set expands super--exponentially on the number of the situations in the tree, making standard types of search algorithms very slow when the number of variables becomes large. 
An exhaustive search or dynamic programming approaches then become unfeasible for all but the smallest of problems and instead a greedy-search algorithm, algorithmic hierarchical clustering (AHC), is typically used. However, AHC still scales poorly and it has been noted by multiple authors \cite{cowell2014causal}, \cite{silander2013dynamic} and \cite{strong22a} that heuristic search methods are needed for model selection with a large numbers of variables. Model selection based on k-means for staged trees has been used but outputs worse MAP estimates than AHC \cite{silander2013dynamic}.

Recently, various papers have suggested \textit{a priori} restricting the search space, for example by using a sub-classes of staged trees such as using cstrees \cite{duarte2021representation}or k-parent staged-trees \cite{leonelli22a}. In this way, model selection becomes feasible and, because of their simplicity, their outputs are often easier to interpret. These approaches has two main drawbacks: firstly, restricting the model space \textit{a priori} restricts the possible independence statements that can be represented; secondly, these subclasses have no obvious extension to classes of staged trees which express asymmetric unfolding of events \cite{shenvi2018modelling}.

Here, we propose an alternative: a novel, heuristic search algorithm that reduces the set of models \textit{a posteriori}. This leads to faster model selection with the potential to learn more complex relationships between variables than in the existing approaches. We then demonstrate our approach's benefits through experiments and a real world example.

\section{Preliminaries}
\subsection{Staged Trees}
An event tree is a directed finite rooted tree with each situation's outgoing edges and corresponding labels representing that situation's potential outcomes. A situation with its children and its induced edges is called a floret. In a staged tree, each situation, $s_i$, has a distribution over its outgoing edges, $\theta_i$. We define two situations -- $s_i, s_j$ -- as being in the same stage if $\theta_i=\theta_j$. The staged tree is a event tree with its staging represented through colours, where two situations are coloured the same if they are in the same stage. We call a staged tree \textit{saturated} if every situation is in a stage by itself. Each possible staging leads to a different staged tree; model selection algorithms aim to provide a good estimate of the MAP staged tree. 

To assign stages to an event tree, we must decide which partitions of situations should be considered as candidate sets i.e. in the context of a given problem, which sets of situations might be plausibly hypothesised to contain exchangeable situations. For example, a staging with situations with a different number of outgoing edges would clearly be erroneous in any context. These conditions are given by a \textit{hyperstage} -- a set of sets called \textit{hypersets}, in which contain the situations that could plausibly be in the same stage. A hyperstage is called a \textit{partitioning hyperstage} if the hypersets form a partition of the set of situations. Typical hyperstages include allowing situations to be in the same stage if they share the same number of outgoing edges and edge labels, and allowing situations be in the same stage if they correspond to the outcome of the same variable.

There exists an alternative graphical representation of a staged tree, known as a Chain Event Graph (CEG) \cite{smith2008conditional}. The CEG provides a more compact representation of the staged tree by combining nodes for which the distributions over all future edges in the tree are the same. 

\subsection{Conjugate learning and model selection}
The posterior probabilities for the staged tree can be obtained through a conjugate updating of a Dirichlet-Multinomial distribution over each situation's edges. For more details, see \cite{CEG_book}.

The model selection algorithm used in this work is based on AHC. In AHC, each situation is initially considered as in a separate stage, and then stages are merged together one at a time by considering which merging would give the largest improvement in model likelihood until there is no possible improvement.

\section{Methods}
\subsection{Totally ordered hyperstage}
To describe our methodology, we introduce the concept of the totally ordered hyperstage.

\begin{definition}[Totally ordered hyperstages]
Suppose a event tree T with a partitioning hyperstage, $\mathbf{H}=\{H_1,H_2, \dots, H_N\}$, and an set of invective ordering functions, $f_n:H_n\to \mathbb{R}$, from each hyperset, $H_n=\{s_{1,n},s_{2,n},\dots,s_{K,n}\}$.  An totally ordered hyperset is $H_n$ with a strict total order induced by $f_n$ such that $s_{i,n}<s_{j,n}$ if $f_n(s_{i,n})<f_n(s_{j,n})$. The totally ordered hyperstage is a set of totally ordered hypersets.\label{OrderedH}
\end{definition}

Now, we can introduce our model selection approach: we use the totally ordered hyperstage to restrict our model space. This is done by preventing stagings of non-consecutive situations in the totally ordered hyperstage. This approach drastically reduces the size of the model space and therefore the time of a search algorithm. When running AHC on a hyperset with $N$ situations, AHC would consider ${N \choose 2}$ possible mergings in the first step. In the worst possible case, AHC would consider $\sum_{n=2}^N{n \choose 2}=\frac{N^3-N}{6}$ possible mergings in its run. In contrast, when running AHC on an totally ordered hyperset with the same number of situations, it considers $N-1$ possible mergings within the first instance, with at worst $\frac{N(N-1)}{2}$ possible mergings considered overall. Therefore, as N increases on the hyperset, the computational time grows cubically. Using the totally ordered hyperset, time grows quadratically, improving its scalability as the number of variables increases. 

\section{Mean posterior Probabilities}
We propose using the mean posterior probability of each situation in the saturated staged tree for each of our ordering functions $f_n$. This function only maps into $\mathbb{R}$ when the maximum number of outgoing edges from any situation is two, restricting this approach to binary trees. 
Definition \ref{OrderedH} also restricts our choice of ordering functions to those that are injective to allow a total ordering. Therefore, the choice of mean posterior probabilities is only suitable if these values are unique. To address this, where there are situations with the same mean transition probability, these situations are automatically placed in the same stage. For example, if multiple situations are given the same prior, they will automatically be placed in the same stage if there are no observations of that situation.  

The mean posterior probability of a situation is the probability of each outcome at a situation. Therefore, choice of mean posterior probability is desirable as it introduces increased interpretability into the model selection process: if two situations are merged together, they must have a comparable probability of their outcomes relative to the set of situations in their hyperset. Note that, from a practical point of view, this prevents the AHC algorithm from combining stages spuriously \cite{non-local} where stages with a large sample size can absorb all stages with small sample size. This choice of ordering function prevents this as the stage with a large sample size would only be able to absorb all of the stages with small sample size if it had comparable mean posterior probability.

Another benefit of this approach compared to existing methods that restrict the space of models for model selection is that this choice of restriction is made \textit{a-posteriori}, meaning that all stagings are possible before any data is collected. Using this method, none of the potential stagings of the staged tree -- and therefore relationships between variables -- are ruled out before the staged tree is fit to data.

We now apply the AHC algorithm using this constraint, calling it the mean posterior clustering (MPC) algorithm.

\subsection{Resize operator}
As stated in Definition \ref{OrderedH}, this approach only works on binary event trees -- event trees where each situation has at most two outgoing edges. However, this is not a major restriction, as any event tree can be written as a binary tree. Research from \cite{gorgen18} and \cite{gorgen21} proves how the statistical equivalence class of staged trees can be traversed through \textit{swap} and \textit{resize} operators and their inverses. Resize operators are of importance here; they work by contracting subtrees in the event tree into a single floret while leaving the rest of the tree invariant. This describes an isomorphism from the set of originally considered staged trees to the set of binary sub-trees. Therefore, each staged tree has an equivalent representation as a staged binary tree: by first transforming a tree into a binary tree, we are simply embedding a search space into a bigger one.

More recently, \cite{hughes22} shows how priors can be set so that they are score-equivalent. Therefore, we can set priors in a way that is consistent between the non-binary and binary tree case for comparison. This is done by setting priors over the root to leave paths; the number of which and how they interact is invariant of the operators used to traverse the statistical equivalence class of staged trees.

Representing any staged tree has the benefit that more complex independence statements can be learnt. For example, when comparing situations in the same hyperstage with three outgoing edges, it is possible that multiple florets have the same distribution over two of their edges but not the third. This relationship could be captured in a binary tree but would be missed in the full tree.

MPC can result in performing more stagings. Suppose we have a hyperstage of a variable with $N$ situations in it. When performing AHC, these $N$ situations need to be staged. When performing MPC on the binary transformation of the tree, the $N$ situations have to be staged $k-1$ times, where $k$ is the number of outgoing edges associated to that variable.

\begin{figure}[h]
    \centering
        \includegraphics[width =0.75\textwidth]{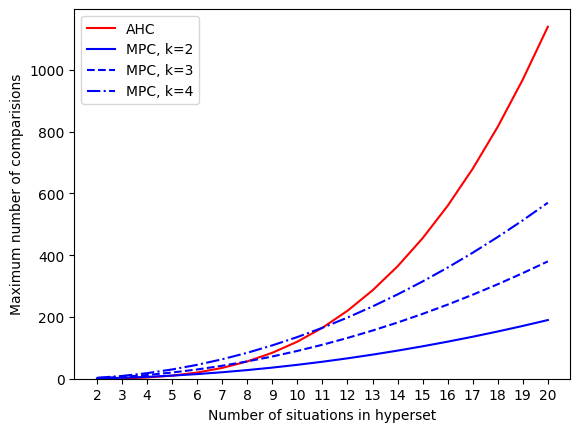}
        \caption{Maximum possible number of considered stagings for differnet model selection algorithms on different trees.}
    \label{timings}
\end{figure}

Therefore, the maximum number of considered stagings in MPC is given by $(k-1)\frac{N(N-1)}{2}$. Figure \ref{timings} shows how this, for different values of $k$, compares to the maximum number of considered stagings in AHC. This shows that for all but the smallest number of situations in the hyperstage, for which the total number of comparisons will the small, the number of comparisons in MPC will be much less than those of AHC. This means that even though when performing MPC more stagings need to take place it still faster than applying AHC.
To represent the tree as a binary tree, we need to have an ordering on the edges to allow the staging to happen. This will be domain-specific depending on the outcomes of interest. This extending of the model space could be done in a way to allow staged trees with structural zero paths to be in the same stage without erroneously treating these as sampling zero paths. 

\section{A Comparative Analysis of Competing Methodologies}
\label{ComparativeAnalysis}
Here, we run comparisons of structural learning algorithms on a number of datasets chosen from existing literature on staged trees. These datasets are available from \cite{JSSv102i06}. We find the MAP estimate using MPC, AHC and AHC on the binary tree. For each dataset, all sampling zero paths were added. We used the score equivalent prior described in \cite{hughes22} to compare staged trees with different underlying event trees. For the purposes of this example, we constrain ourselves to using a fixed parameter $\alpha$ to the number of leaves of each tree -- setting a uniform prior over the leaves of the tree for a model that sets all situations in distinct stages -- rather than select over different values of alpha. The hyperstage was set so that situations relating to the same variable could be in the same stage.

The event trees were made binary using the following process: for florets with more than two outgoing edges selecting outcomes, by order of appearance in the data-frame, create two florets. One provides the selected outcome on one edge and the other edge leads to a floret with the rest of the outcomes. This was performed iteratively until the resulting tree was binary. For an example see Figure \ref{resize}.

\begin{table}[]
\centering
\begin{adjustbox}{width=1.2\textwidth,center=\textwidth}
\small
\begin{tabular}{cr|rr|rr|rr}
 &
  \multicolumn{1}{c|}{} &
  \multicolumn{2}{c|}{MPC} &
  \multicolumn{2}{c|}{AHC binary tree} &
  \multicolumn{2}{c}{AHC} \\
Dataset &
  \multicolumn{1}{c|}{\# Leaves} &
  \multicolumn{1}{c}{Time(s)} &
  \multicolumn{1}{c|}{Log-likelihood} &
  \multicolumn{1}{c}{Time(s)} &
  \multicolumn{1}{c|}{Log-likelihood} &
  \multicolumn{1}{c}{Time(s)} &
  \multicolumn{1}{c}{Log-likelihood} \\ \hline
Asym &  16 &  \textbf{0.01} &  \textbf{-2411.22} &  * &  * &  0.03 &  -2423.67 \\
Pokemon &  32 &  \textbf{0.25} &  \textbf{-3251.94} &  * &  * &  1.26 &  \textbf{-3251.94} \\
Titanic &  32 &  \textbf{0.15} &  \textbf{-5210.51} &  0.27 &  \textbf{-5210.51} &  1.05 &  -5243.58 \\
reinis &  64 &  \textbf{0.92} &  \textbf{-6712.44} &  * &  * &  9.30 &  -6715.51 \\
PhDArticles &  144 &  \textbf{5.17}  & -4118.96 &  87.61 &  \textbf{-4118.72} &  207.53 &  -4198.83 \\
chestSim50000 &  256 &  \textbf{4.56} &  \textbf{-112446.73} &  * &  * &  2981.84 &  -113458.87 \\
monks1 &  432 &  \textbf{309.70} &  \textbf{-2625.38} &  - &  - &  - &  -
\end{tabular}
\end{adjustbox}
\caption{Results showing the outcomes of the experiments. Smallest time and largest log-likelihood are in bold. An asterisks (*) is used to show when the original tree was binary. A hyphen (-) shows when the experiment timed out and took longer than 10,000 seconds. Experiments were performed on a laptop with 16GB of RAM with 4core i7 2.6ghz cpu.}
\label{Table_compariosn}
\end{table}

Table \ref{Table_compariosn} shows the results of our comparison. MPC was the fastest model selection algorithm for all of the datasets considered, with it being orders of magnitude faster than AHC on the larger datasets. Regarding log-likelihoods, the binary event trees were larger than the original event tree. MPC often outperformed AHC. This illustrates that, although in the binary case it may consider far less partitions than AHC does, MPC can achieve similar, if not better, performance than AHC in a much faster time.

\section{Christchurch Health and Development Study Example}
We next look at a dataset that has previously been studied in staged tree literature \cite{BN_to_CEG} \cite{cowell2014causal} and apply our methodology. This dataset comes from the Christchurch Health and Development Study (CHDS), as detailed in \cite{BN_to_CEG}. This is a longitudinal cohort study, taking place over 30 years, of 1265 children born in mid-1977 in Christchurch, New Zealand. As in \cite{BN_to_CEG} and \cite{cowell2014causal}, we are interested in the following 4 discrete variables for 890 children for which complete data was available:

\begin{itemize}
    \item $X_S$ = family social background (High, Low), to henceforth be known as "social background"
    \item $X_E$ = family economic situation (High, Low), to henceforth be known as "economic situation"
    \item $X_H$ = child hospital admission, (Yes, No)
    \item $X_L$ = number of family life events, such as death of close relatives or divorce (Low, Average, High)
\end{itemize}

Previous work on this dataset in \cite{BN_to_CEG} has shown how a staged tree can outperform a Bayesian network on this dataset; \cite{barclay_hutton_smith_2014} showed the above ordering is that which gives the highest scoring staged tree. For our methodology, we must first perform a \textit{resize} on the floret associated with number of family life events $X_L$ to make the tree binary.

\begin{figure}[h]
\centering
\begin{subfigure}[b]{0.25\textwidth}
\includegraphics[width=\textwidth]{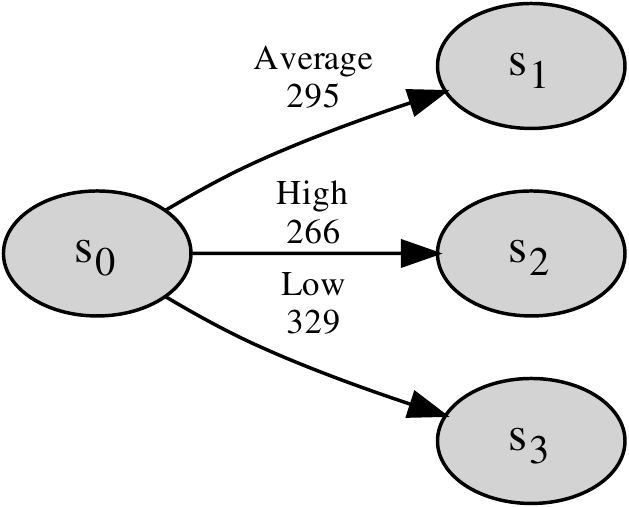} 
\caption{Floret $X_L$}
\label{floret}
\end{subfigure}
\hfill
\begin{subfigure}[b]{0.45\textwidth}
    \includegraphics[width=\textwidth]{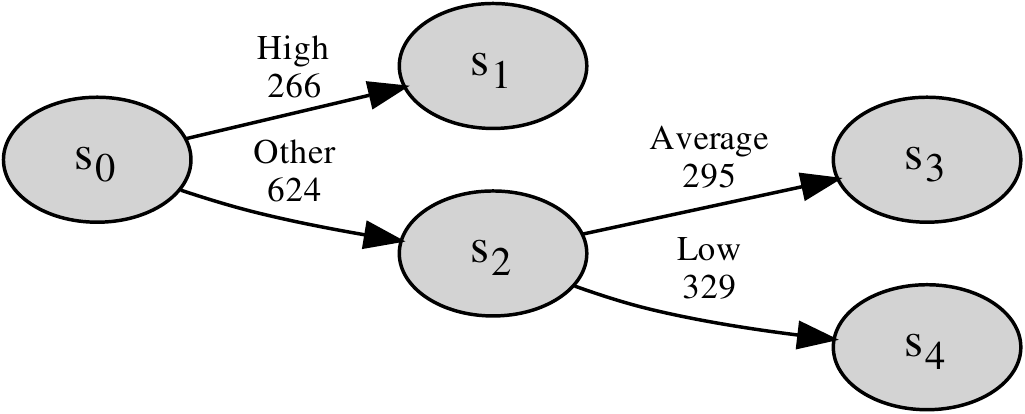}
\caption{Resize of $X_L$}
\label{b_floret}
\end{subfigure}
\caption{Resizing of the floret $X_L$ so that the event tree is binary.}
\label{resize}
\end{figure}

We split up the variable of life events into two binary florets:
\begin{itemize}
    \item $X_{L-high}$: Was the number of life events High? (High, Other)
    \item $X_{L-average/low}$: If "other", was the number of life events Average or low? (Average, Low).
\end{itemize}
This resizing is shown in Figures \ref{floret} and \ref{b_floret}.  Using the binary florets gives the event tree in Figure \ref{bin_tree}. This embedding enables us to focus more closely on households that have a high number of life events.

The model selection in this example were set up in the same way as described in Section \ref{ComparativeAnalysis}. Running MPC on the binary tree gives the same output as running AHC on the binary tree. Both binary tree outputs outperform running AHC on the original tree. We also looked at the other two ways of making this tree binary -- by resizing for average and low number of life events. Resizing for high numbers of life events gave the best log-likelihood, although all three models outperformed the output obtained by AHC on the original tree.

\begin{figure}[h]
    \centering
        \includegraphics[width =\textwidth]{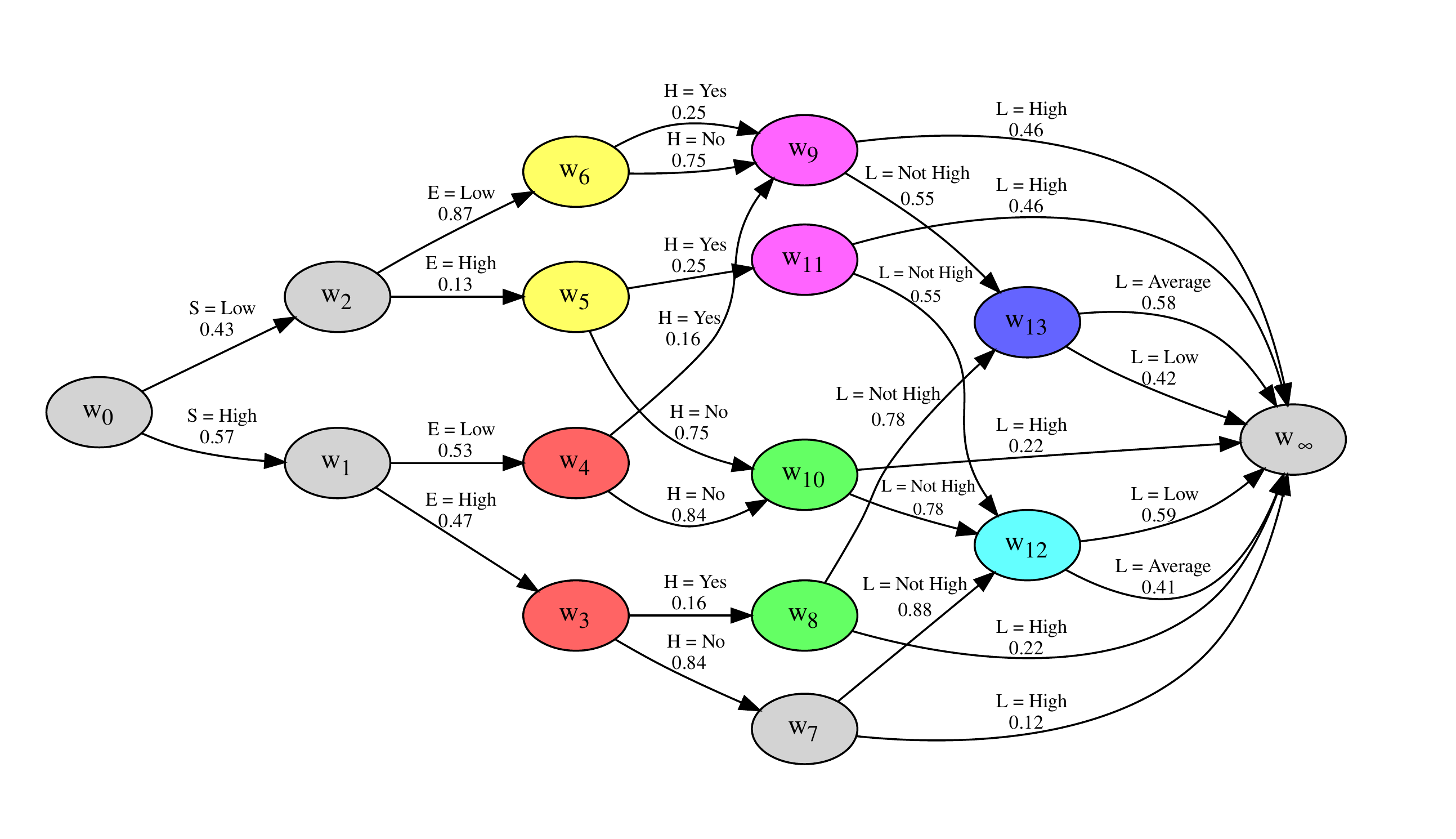}
        \caption{CEG showing data from the Christchurch Health and Development Study. S: social background; E: economic situation; H: admitted to hospital; L: number of life events.}
    \label{CEG}
\end{figure}

It is first important to note that variables that are the same in the binary and non-binary trees -- $X_S$, $X_E$ and $X_H$ -- the staging is unsurprisingly the same, as the staging of this part of the staged tree is unaffected by the resizing. 

However, for the part of the event tree that has been resized, $X_L$, new stage structure has been learned that is not present in the non-binary MAP staged tree. Figure \ref{CEG} has the situations corresponding to the variable $X_{L-high}$ in three stages. These are ordered below with increasing probability of a high number of life events:
\begin{itemize}
    \item  There exists a single situation for individuals who have high social background and economic situation and were not admitted to hospital ($w7$). These individuals have the smallest probability of a high number of life events.
    \item  Individuals that have low social background, high economic situation and have not been admitted to hospital ($w10$), individuals that have high social background, low economic situation and have not been admitted to hospital ($w10$) and individuals with high social background and economic situation who have been admitted to hospital ($w8$) have the same probability of high numbers of life events.
    \item Individuals that have low social background and economic situation (irrespective of hospital admission) ($w9$), individuals who have high social background but low economic situation and have been admitted to hospital ($w9$) and individuals with low social background but high economic situation who have been admitted to hospital ($w11$) all have the highest probability of a high number of life events.

\end{itemize}

This staging may appear complex at first glance but there exists an intuitive rationale behind the learned staging. Our model shows that low social background, low economic situation and being admitted to hospital all have the potential to increase the chance of having a high number of life events. Having none of these leads to a staging that has the smallest mean posterior probability of a high number of life events. Being admitted to hospital (unless you have high social status and economic situation) or having low social status or economic background is linked to having many more life events. The staging of $X_{L-average/low}$, where an individual does not have a high number of life events, does not follow the same pattern of specific variable outcomes increasing the risk of a higher number of life events. This shows that by expanding the model space, we can learn further insights from the data, not previously possible, as the staging obtained by MPC is not statistically equivalent to any staging on the non-binary tree. 

\section{Discussion}
The MPC algorithm for model selection across staged trees is a novel structural search algorithm which outputs similarly- or better-performing models than the traditional AHC algorithm at a lower computational cost, because MPC scales quadratically rather than cubically. Through experiments and an example, we have illustrated the benefits MPC can provide for rapid computation and model accuracy.
We have also defined a new class of staged trees: the binary staged tree. Binary staged trees provide increased explainability through expanding the model space allowing for better-fitting models than non-binary staged trees, irrespective of whether AHC or MPC is used. We have illustrated the benefits of binary staged trees through experiments and an example, which identifies new independence statements not found in the non-binary tree. 
There is scope for further investigation into both the MPC algorithm and binary trees. One avenue for further research is to explore the suitability of choices other than the mean posterior probability for ordering functions on the hyperstage. 

\bibliography{cite}

\appendix
\section{Event tree of Christchurch Health and Development Study Example}

\begin{figure}[h]
    \centering
        \includegraphics[width =0.8\textwidth]{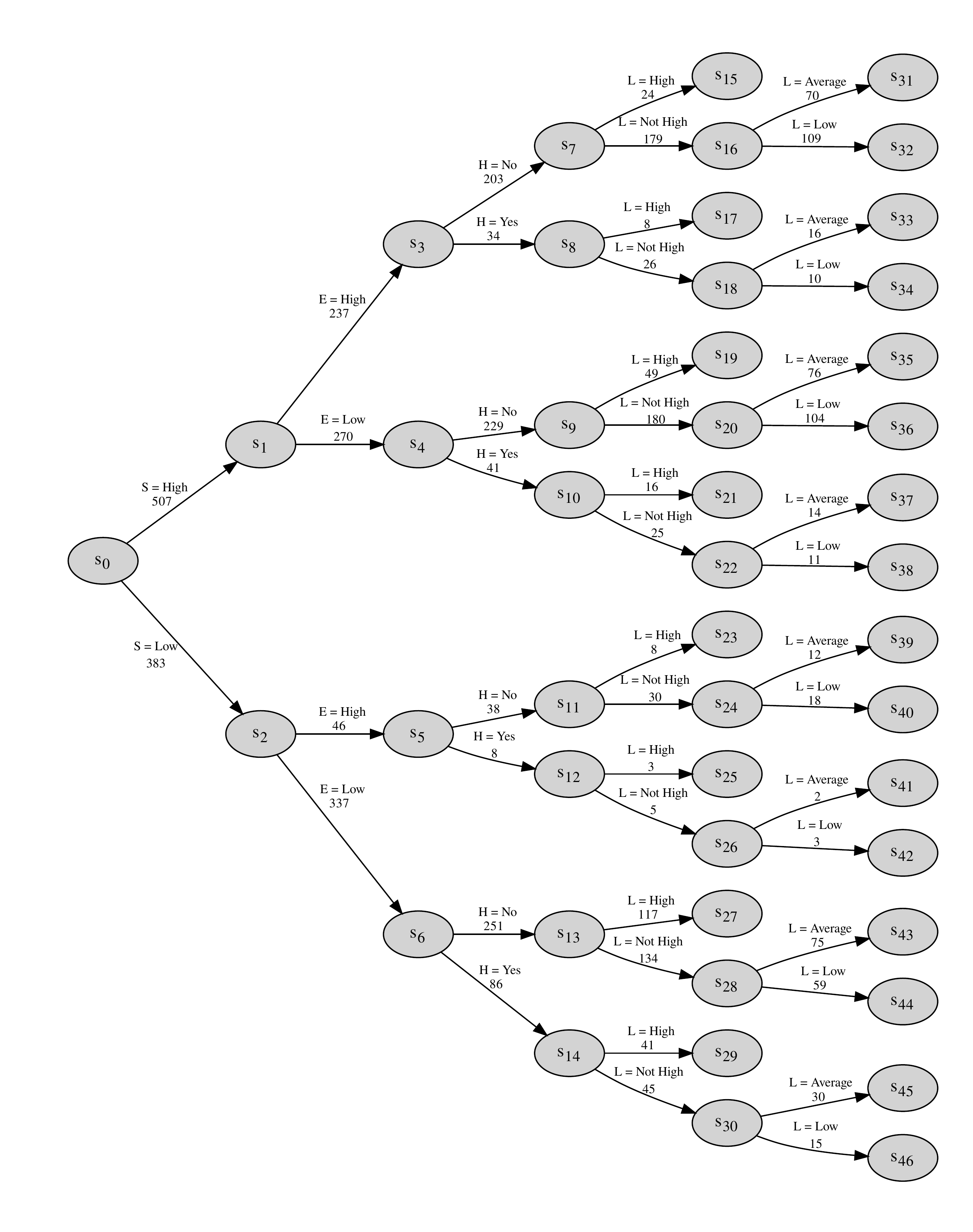}
        \caption{Binary event tree from the Christchurch Health and Development Study. S: social background; E: economic situation; H: admitted to hospital; L: number of life events.}
    \label{bin_tree}
\end{figure}

\end{document}